\documentclass{rmf-d}
\usepackage{nopageno,rmfbib,multicol,times,epsf,amsmath,amssymb,cite}
\usepackage[latin1]{inputenc}
\usepackage[]{caption2}
\usepackage{graphicx}
\usepackage{booktabs}

\usepackage{hyperref}

\hypersetup{%
colorlinks,%
citecolor=blue,urlcolor=blue,linkcolor=blue,%
hypertexnames=false,%
pdftitle={Three-pion scattering in Chiral Perturbation Theory},%
pdfauthor={T. Husek}%
}

\allowdisplaybreaks

\def\rmfcornisa{\hfill LU TP 22-22}

\def\rmfcintilla{}
\clearpage \rmfcaptionstyle 
\begin{document}
\title{Three-pion scattering in Chiral Perturbation Theory
\vspace{-6pt}}
\author{Tom\'a\v{s} Husek}
\address{Department of Astronomy and Theoretical Physics, Lund University, Box 43, SE-22100 Lund, Sweden}
\maketitle
\begin{abstract}
\vspace{1em}
We present the results on the relativistic six-pion scattering amplitude at low energy, calculated at $\mathcal{O}(p^4)$ within the framework of the massive $\text{O}(N)$ nonlinear sigma model extended to the next-to-leading order in the chiral counting.
For $N=3$, this approach corresponds to the two(-quark)-flavor Chiral Perturbation Theory.
We also present the expressions for the pion mass, pion decay constant and the four-pion amplitude in the case of $N$ (meson) flavors at $\mathcal{O}(p^4)$.
\vspace{1em}
\end{abstract}
\keys{  Chiral perturbation theory, pions, hadron-hadron interactions, scattering amplitudes \vspace{-4pt}}
\pacs{  \bf{\textit{
12.39.Fe Chiral Lagrangians,
11.30.Rd Chiral symmetries,
14.40.Aq pi, K, and eta mesons
 }}    \vspace{-4pt}}
\begin{multicols}{2}

\section{Introduction}

To study interactions of hadrons perturbatively in the low-energy region, we cannot directly employ Quantum Chromodynamics (QCD), the fundamental theory of strong interactions.
This is due to confinement, stemming from the underlying non-abelian structure of QCD.
Instead, we are left with using alternative approaches, for instance, Chiral Perturbation Theory (ChPT)~\cite{Weinberg:1978kz,Gasser:1983yg}.
Many observables have been calculated applying this very successful effective field theory to a high loop order.
However, this is not the case for the six-pion amplitude, which has been until now only known at tree level~\cite{Osborn:1969ku,Low:2019ynd,Bijnens:2019eze}.
Since it has been recently estimated using lattice QCD~\cite{Mai:2018djl,Blanton:2019vdk,Mai:2019fba,Culver:2019vvu,Fischer:2020jzp,Hansen:2020otl,Brett:2021wyd,Blanton:2021llb}, it seems interesting and complementary to provide a consistent NLO calculation of the six-pion amplitude at one-loop level in ChPT~\cite{Bijnens:2021hpq}.

\section{Theoretical setting}

For the calculation which follows, we used a simple generalization of two(-quark)-flavor ChPT --- the massive O($N+1$)/O($N$) nonlinear sigma model extended to the next-to-leading order (NLO) in the chiral counting, taking thus into account $N$ meson (pion) flavors --- the results of which it reproduces for $N=3$:
\begin{equation}
\begin{split}
\mathcal{L}
&=\frac{F^2}2\,\partial_\mu\Phi^\mathsf{T}\partial^\mu\Phi+F^2\chi^\mathsf{T}\Phi\\
&+l_1\big(\partial_\mu\Phi^\mathsf{T}\partial^\mu\Phi\big)\big(\partial_\nu\Phi^\mathsf{T}\partial^\nu\Phi\big)\\
&+l_2\big(\partial_\mu\Phi^\mathsf{T}\partial_\nu\Phi\big)\big(\partial^\mu\Phi^\mathsf{T}\partial^\nu\Phi\big)\\
&+l_3\big(\chi^\mathsf{T}\Phi\big)^2
+l_4\partial_\mu\chi^\mathsf{T}\partial^\mu\Phi\,.
\end{split}
\label{eq:L}
\end{equation}
Above, $\Phi$ is a real vector of $N+1$ components which satisfies $\Phi^\mathsf{T}\Phi=1$, and $\chi^\mathsf{T}=\left(M^2,\,\vec 0\,\right)$.
At the leading order (LO), we have two parameters: pion decay constant $F$ and mass $M$; at NLO, four additional monomials relevant for our applications show up, accompanied with low-energy constants $l_i$.
These carry both ultraviolet (UV)-divergent and convergent parts, the latter of which are free parameters in the theory and need to be extracted from experiment or lattice QCD.
We write
\begin{equation}
\begin{split}
l_i
&=-\kappa\,\frac{\gamma_i}2\frac1{\tilde\epsilon}+l_i^\text{r}\,,\quad\kappa\equiv\frac1{16\pi^2}\,,\\
\frac1{\tilde\epsilon}
&\equiv\frac{1}{\epsilon}-\gamma_\text{E}+\log4\pi-\log\mu^2+1\,,\quad
\epsilon=2-d/2\,.
\end{split}
\label{eq:li2}
\end{equation}
The divergent parts are uniquely fixed from studying the pion mass, decay constant and four-pion amplitude at NLO.
We find
\begin{align}
\gamma_1&=\frac N2-\frac76\,,\quad&
\gamma_3&=1-\frac N2\,,\\
\gamma_2&=\frac23\,,\quad&
\gamma_4&=N-1\,.
\end{align}

To work with the Lagrangian~\eqref{eq:L}, one needs to expand it in terms of pion fields $\phi_i$.
This can be done with the use of a particular parameterization for fields $\Phi$, the whole class of which can be written in general as
\begin{equation}
\Phi
=\left(\sqrt{1-\varphi\,f^2(\varphi)},\,f(\varphi)\,\frac{\pmb{\phi}^\mathsf{T}}F\right)^\mathsf{T}\,.
\end{equation}
Above, $\varphi\equiv\frac{\pmb{\phi}^\mathsf{T}\pmb{\phi}}{F^2}$, with $\pmb{\phi}^\mathsf{T}=(\phi_1,\dots,\phi_N)$ being a real vector of $N$ components (flavors), and $f(x)$ is any analytical function satisfying $f(0)=1$.
For a practical calculation, it is convenient to employ more than one parameterization, utilizing, as a neat cross-check, the fact that the physical amplitudes should be parameterization-independent.

\section{Four-pion amplitude}

We start with the four-pion amplitude which turns out to be an important ingredient for the six-pion amplitude.
As is fairly well-known, the four-pion amplitude can be written, due to its invariance under rotation in the isospin space and crossing symmetry, as (all pion four-momenta $p_i$ incoming, flavors $f_i$)
\begin{equation}
\begin{split}
&A_{4\pi}(p_1,f_1,p_2,f_2,p_3,f_3,f_4)\\
&=\delta_{f_1f_2}\delta_{f_3f_4}A(p_1,p_2,p_3)\\
  &+\delta_{f_1f_3}\delta_{f_2f_4}A(p_3,p_1,p_2)\\
  &+\delta_{f_2f_3}\delta_{f_1f_4}A(p_2,p_3,p_1)\,,
\end{split}
\end{equation}
separating thus the flavor structure from the momentum-dependent part given in terms of a single (four-pion) sub-amplitude $A(s,t,u)=A(p_1,p_2,p_3)$.
In the last expression, we introduced the standard Mandelstam variables $s=(p_1+p_2)^2$, $t=(p_1+p_3)^2$, $u=(p_2+p_3)^2$, which satisfy the on-shell relation $s+t+u=4M^2$.

At LO, the amplitude stems from a single tree-level Feynman diagram with the text-book result $A^{(2)}(s,t,u)=\left(s-M_\pi^2\right)/F_\pi^2$.
At NLO, we have two topologies of in total four loop diagrams and a counter-term.
These, together with the wave-function renormalization and the NLO expressions for the pion mass and decay constant,
\begin{equation}
\begin{split}
M^2&=M_\pi^2-\frac{M_\pi^4}{F_\pi^2}\left[2l_3^\text{r}+\frac12(N-2)L\right],\\
\frac1{F^2}&=\frac1{F_\pi^2}\left\{1+2\,\frac{M_\pi^2}{F_\pi^2}\left[l_4^\text{r}-\frac12(N-1)L\right]\right\},
\end{split}
\end{equation}
give us the final expression for the parameterization-independent and UV-finite four-pion amplitude:
\begin{align}
&F_\pi^4A^{(4)}(s,t,u)
=(t-u)^2\left(
    -\frac5{36}\,\kappa
    -\frac16\,L
    +\frac12\,l_2^\text{r}
\right)\notag\\
&+s^2\bigg[
    \bigg(\frac{11}{12}-\frac N2\bigg)\kappa
    +\bigg(1-\frac N2\bigg)L
    +2l_1^\text{r}
    +\frac12\,l_2^\text{r}
\bigg]\notag\\
&+M_\pi^2s\bigg[
    \bigg(N-\frac{29}9\bigg)\kappa
    +\bigg(N-\frac{11}3\bigg)L
    -8l_1^\text{r}
    +2l_4^\text{r}
\bigg]\notag\\
&+M_\pi^4 \bigg[
    \bigg(\frac{20}{9}-\frac N2\bigg)\kappa
    +\bigg(\frac83-\frac N2\bigg)L
    +8\,l_1^\text{r}
    +2l_3^\text{r}
    -2l_4^\text{r}
\bigg]\notag\\
&+\bar J(s)\bigg[
    \bigg(\frac N2-1\bigg) s^2
    + (3-N) M_\pi^2 s
    + \bigg(\frac N2-2\bigg) M_\pi^4
\bigg]\notag\\
&+\bigg\{\frac16\,\bar J(t)\big[
    2 t^2
    - 10 M_\pi^2 t
    - 4 M_\pi^2 s
    + s t
    + 14 M_\pi^4
\big]\notag\\
&+(t\leftrightarrow u)\bigg\}\,.
\label{eq:A4pisub_NLO}
\end{align}
It depends explicitly on the number of meson flavors $N$ and it is consistent with the previous results found in literature~\cite{Gasser:1983yg,Dobado:1994fd,Bijnens:2009zi,Bijnens:2010xg}.
This exact form will be used later on and it is a generalization of the results shown in Refs.~\cite{Bijnens:1995yn,Bijnens:1997vq} beyond $N=3$.
Above, we have used
\begin{equation}
L\equiv\kappa\log\frac{M_\pi^2}{\mu^2}\,.
\end{equation}

\section{Six-pion amplitude}

Compared to the four-pion amplitude, the combinatorics becomes significantly more involved in the case of the six-pion amplitude.
In the preceding section and considering four pions, we had only three channels (permutations) or ways how to distribute four pions in two pairs.
The six-pion amplitude, already at LO (see Fig.~\hyperlink{fig:I}{I}), is represented by two topologies of Feynman diagrams, which are related to two sets of permutations: There are ten ways how to distribute six pions in two groups of three ($P_{10}$; relevant for one-particle-reducible (1PR) topologies) and fifteen ways to distribute them in three pairs ($P_{15}$; relevant for the six-pion subamplitude discussed later).
It becomes natural that we write the complete six-pion amplitude as a sum of two pieces:
\begin{equation}
A_{6\pi}
=A_{6\pi}^{(4\pi)}+A_{6\pi}^{(6\pi)}\,.
\end{equation}
The first piece can be written in terms of the four-pion amplitude, while the second part is the remainder.
%
\vspace{2mm}
\vspace{2mm}
\hrule\vspace{0.5mm}\hrule
\vspace{2mm}
{\noindent\strut\small
\sc Figure \hypertarget{fig:I}{I}.\ {\rm
    Six-pion amplitude at the leading order.
    The multiplicities of the respective topologies based on all the possible permutations of the external legs are quoted.
}}\\
\begin{minipage}{\columnwidth}
\vspace{4mm}
\centering
\begin{minipage}[b]{0.45\columnwidth}
\centering
\includegraphics[width=0.75\columnwidth]{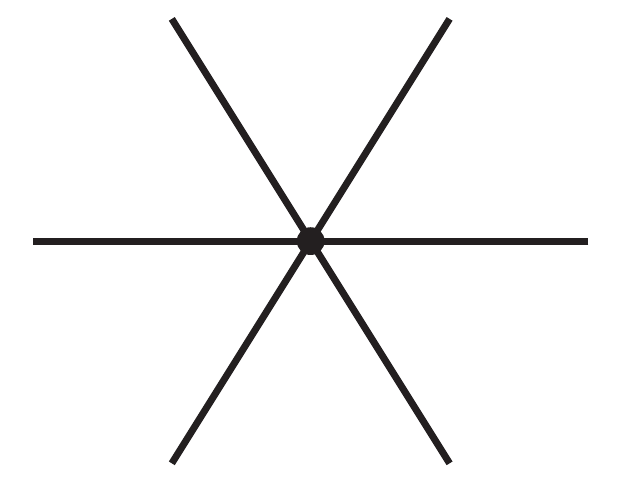}\\
{\noindent\strut\small (a)\ {\rm 1$\times$}}
\end{minipage}
\begin{minipage}[b]{0.5\columnwidth}
\centering
\includegraphics[width=0.85\columnwidth]{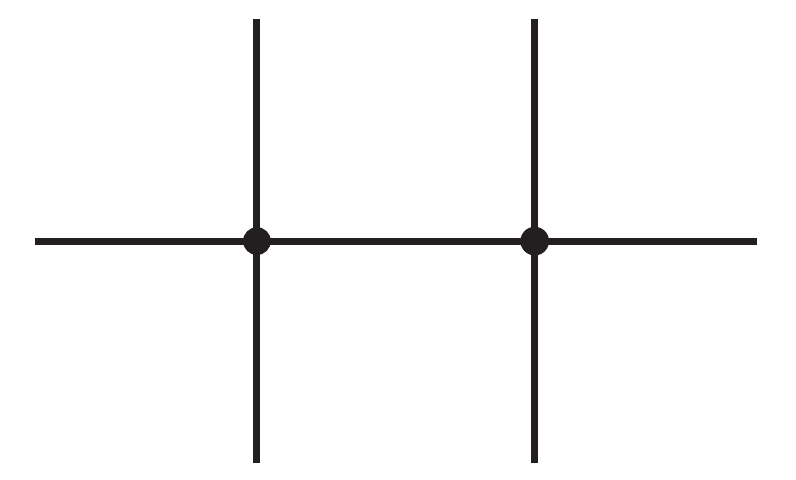}\\
{\noindent\strut\small (b)\ {\rm 10$\times$}}
\end{minipage}
\vspace{2mm}
\end{minipage}
%
\vspace{2mm}
\hrule\vspace{0.5mm}\hrule
\vspace{2mm}
{\noindent\strut\small
\sc Figure \hypertarget{fig:II}{II}.\ {\rm
    Six-pion-amplitude topologies at the next-to-leading order.
    We again quote the multiplicities of the respective diagrams.
}}\\
\begin{minipage}{\columnwidth}
\vspace{4mm}
\centering
\begin{minipage}[t]{0.33\columnwidth}
\centering
\includegraphics[width=0.9\columnwidth]{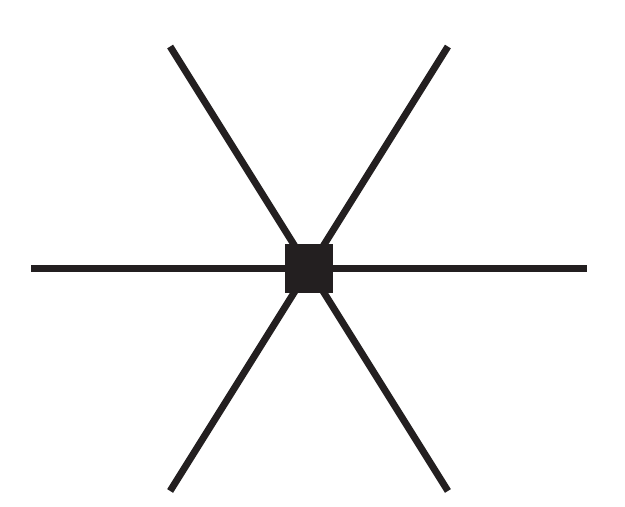}\\
{\noindent\strut\small (a)\ {\rm 1$\times$}}
\end{minipage}
\begin{minipage}[t]{0.32\columnwidth}
\centering
\includegraphics[width=0.9\columnwidth]{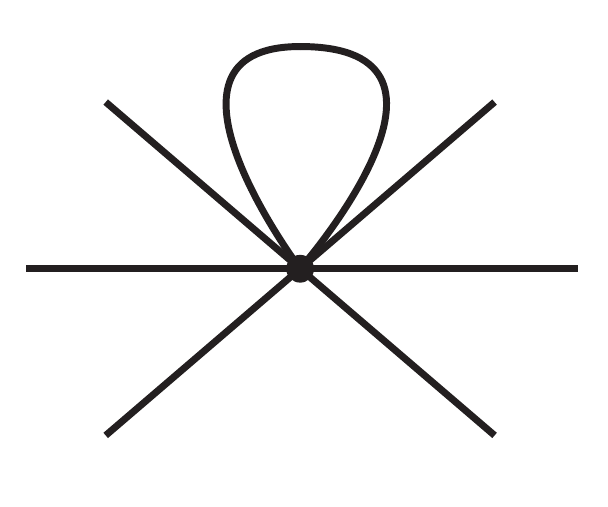}\\
{\noindent\strut\small (b)\ {\rm 1$\times$}}
\end{minipage}
\begin{minipage}[t]{0.32\columnwidth}
\centering
\includegraphics[width=0.9\columnwidth]{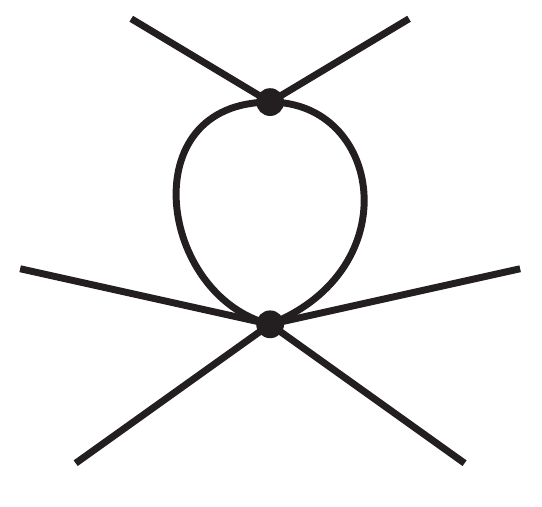}\\
{\noindent\strut\small (c)\ {\rm 15$\times$}}
\end{minipage}

\vspace{2mm}
\begin{minipage}[t]{0.33\columnwidth}
\centering
\includegraphics[width=0.9\columnwidth]{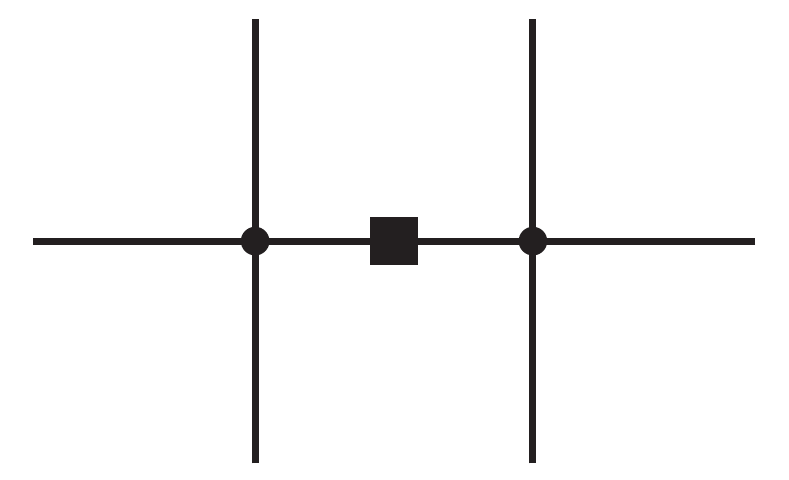}\\
{\noindent\strut\small (d)\ {\rm 10$\times$}}
\end{minipage}
\begin{minipage}[t]{0.32\columnwidth}
\centering
\includegraphics[width=0.9\columnwidth]{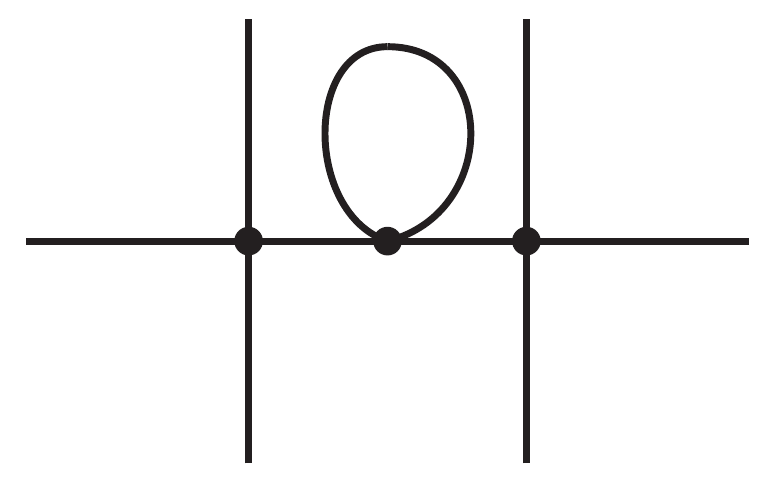}\\
{\noindent\strut\small (e)\ {\rm 10$\times$}}
\end{minipage}
\begin{minipage}[t]{0.32\columnwidth}
\centering
\includegraphics[width=0.9\columnwidth]{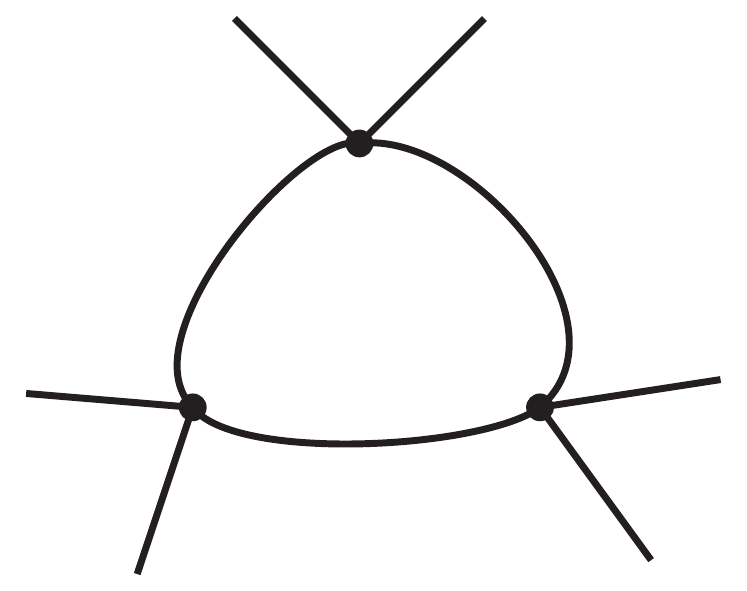}\\
{\noindent\strut\small (f)\ {\rm 15$\times$}}
\end{minipage}

\vspace{2mm}
\begin{minipage}[t]{0.33\columnwidth}
\centering
\includegraphics[width=0.9\columnwidth]{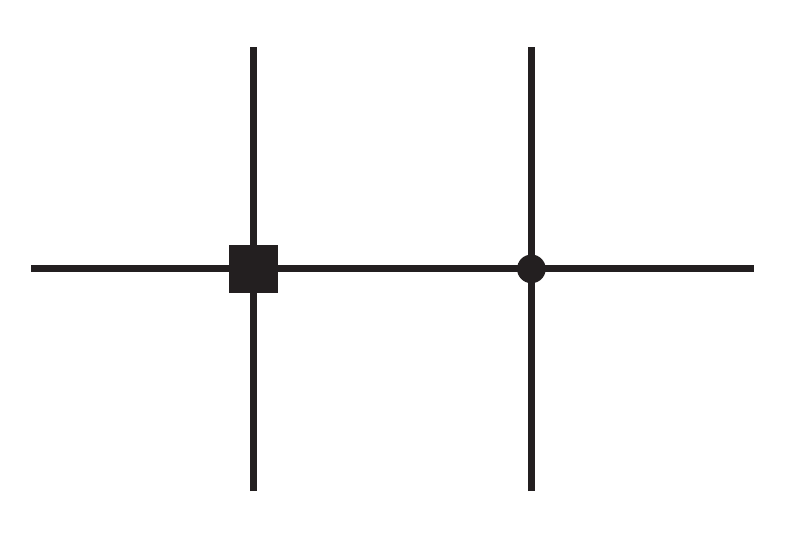}\\
{\noindent\strut\small (g)\ {\rm 20$\times$}}
\end{minipage}
\begin{minipage}[t]{0.32\columnwidth}
\centering
\includegraphics[width=0.9\columnwidth]{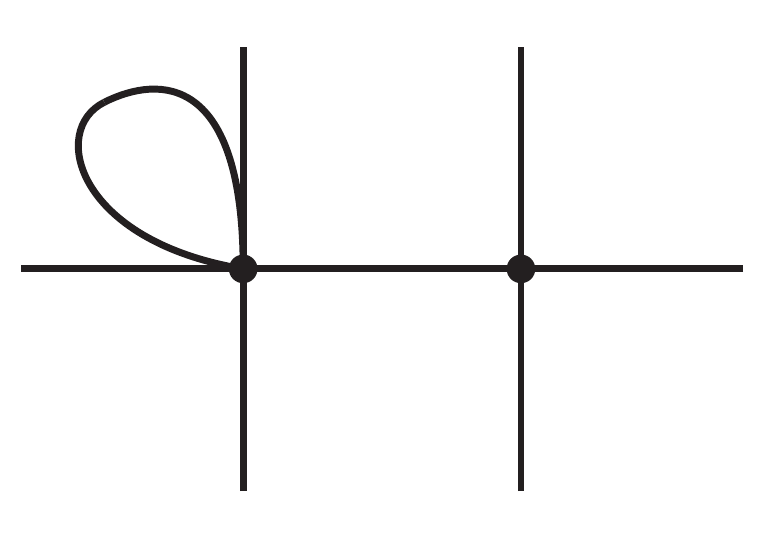}\\
{\noindent\strut\small (h)\ {\rm 20$\times$}}
\end{minipage}
\begin{minipage}[t]{0.32\columnwidth}
\centering
\includegraphics[width=0.9\columnwidth]{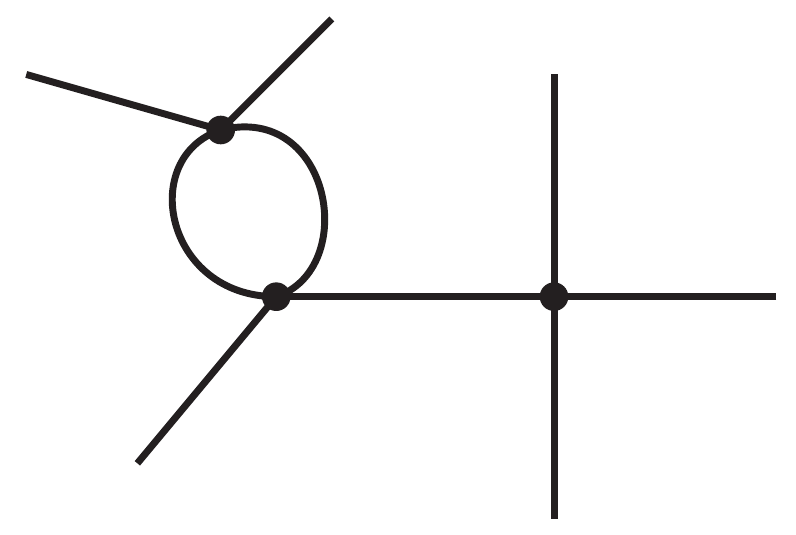}\\
{\noindent\strut\small (i)\ {\rm 60$\times$}}
\end{minipage}
\vspace{4mm}
\end{minipage}
%

This pattern also holds at NLO (see Fig.~\hyperlink{fig:II}{II}).
We have tadpole [\hyperlink{fig:II}{II}(b)], bubble [\hyperlink{fig:II}{II}(c)] and triangle [\hyperlink{fig:II}{II}(f)] diagrams contributing together with the counter-term contributions [\hyperlink{fig:II}{II}(a)] to the one-particle-irreducible (1PI) part, and then pole diagrams [\hyperlink{fig:II}{II}(d, e, g--i)], which, when studied in detail, combine together into two NLO four-pion amplitudes connected with an off-shell leg (flavor $f_\text{o}$):
\begin{equation}
\begin{split}
A_{6\pi}^{(4\pi)}
&\equiv\sum_{P_{10},\,f_\text{o}}
A_{4\pi}(p_i,f_i,p_j,f_j,p_k,f_k,f_\text{o})\\
&\times\frac{(-1)}{p_{ijk}^2-M^2_\pi}\,
A_{4\pi}(p_l,f_l,p_m,f_m,p_n,f_n,f_\text{o})\,.
\end{split}
\label{eq:A6pi4pi}
\end{equation}

The four-pion amplitude can be decomposed in the same way as before in terms of one momentum-dependent subamplitude $A(s,t,u)$, where now the Mandelstam variables satisfy the off-shell relation $s+t+u=3M^2_\pi+p_{ijk}^2$, with $p_{ijk}\equiv p_i+p_j+p_k$.

The residue of the expression in Eq.~\eqref{eq:A6pi4pi} is unique:
When the propagator goes on-shell, the four-pion amplitudes also become on-shell.
And we already know that those are unique.
We are then left with freedom for the off-shell extrapolation, while the choice-dependent remainder is deferred to the 1PI part of the six-pion amplitude.
For our expressions, we chose to use the particular form of the four-pion subamplitude given in Eq.~\eqref{eq:A4pisub_NLO}.

The 1PI part of the six-pion amplitude can be written as
\begin{equation}
A_{6\pi}^{(6\pi)}
\equiv\sum_{P_{15}}\delta_{f_if_j}\delta_{f_kf_l}\delta_{f_mf_n}A(p_i,p_j,p_k,p_l,p_m,p_n)\,,
\label{eq:A6piP15}
\end{equation}
separating thus again the flavor structure from a single momentum-dependent subamplitude denoted above as $A(p_1,p_2,p_3,p_4,p_5,p_6)$.
Since the pole structure is already reproduced by the 1PR part, the real part of the six-pion subamplitude does not contain any poles; however, the imaginary part of the triangle one-loop integrals can contain poles.
$A(p_1,p_2,p_3,p_4,p_5,p_6)$ is a function of three pairs of momenta, being fully symmetric under the interchange of any of the pairs as well as of the momenta within each pair.

At LO, we have a simple expression
\begin{equation}
\begin{split}
A^{(2)}
&\equiv A^{(2)}(p_1,p_2,p_3,p_4,p_5,p_6)\\
&=\frac{1}{F_\pi^4}\left(2p_1\cdot p_2+2p_3\cdot p_4+2p_5\cdot p_6+3M_\pi^2\right),
\end{split}
\label{eq:defA2}
\end{equation}
the form of which, regarding the dependence on momenta, is the only one consistent with the symmetries stated above at the given order.
Finally, our main result is the NLO six-pion subamplitude.
We write it in terms of many parts:
\begin{equation}
\begin{split}
&F_\pi^6 A^{(4)}
\equiv F_\pi^6 A^{(4)}(p_1,p_2,p_3,p_4,p_5,p_6)\\
&=A_J^{(1)}+A_J^{(2)}+A_\pi+A_L+A_l\\
&+A_{C_3}+A_{C_{21}}^{(1)}+A_{C_{21}}^{(2)}+A_{C_{11}}+A_C^{(1)}+A_C^{(2)}+A_C^{(3)}\,.
\end{split}
\label{eq:defA4}
\end{equation}
The pieces listed in the last row are related to tensor triangle one-loop integrals; the reduction to the scalar one-loop integrals would lead to enormous expressions given the number of kinematic invariants that are present.
We have instead chosen a specific redundant basis of integrals that have good symmetry properties and allow us to present the results in a rather compact way.
All the subparts of the six-pion subamplitude from Eq.~\eqref{eq:defA4} then have the correct symmetry properties.
The expressions for these can be found in Ref.~\cite{Bijnens:2021hpq}.

\section{Results}

To present numerical results, we need to adopt a particular kinematical setting to reduce the number of relevant variables.
We choose a symmetric $3\to3$ scattering configuration in which all the pions have the same momentum (modulus $p$) and consequently the energy $E_p=\sqrt{M_\pi^2+p^2}$. The suitable four-momenta are then
\begin{align}
\label{eq:kinematics}
    p_1&=\left(E_p,p,0,0\right),\notag\\
    p_2&=\left(E_p,-\frac{1}{2}p,\frac{\sqrt{3}}{2}p,0\right),\notag\\
    p_3&=\left(E_p,-\frac{1}{2}p,-\frac{\sqrt{3}}{2}p,0\right),\notag\\
    p_4&=\left(-E_p,0,0,p\right),\notag\\
    p_5&=\left(-E_p,\frac{\sqrt{3}}{2}p,0,-\frac{1}{2}p\right),\notag\\
    p_6&=\left(-E_p,-\frac{\sqrt{3}}{2}p,0,-\frac{1}{2}p\right).
\end{align}
We use the following numerical inputs~\cite{Bijnens:2014lea,Colangelo:2001df,Aoki:2016frl}:
\begin{align}
    M_\pi&=0.139570\,\text{GeV}\,,\quad&  \bar l_1&=-0.4\,,\notag\\
    F_\pi&=0.0927\,\text{GeV}\,,&  \bar l_2&=4.3\,,\notag\\
    \mu&=0.77\,\text{GeV}\,,&  \bar l_3&=3.41\,,\notag\\
    N&=3\,,&  \bar l_4&=4.51\,.
\end{align}

In Fig.~\hyperlink{fig:III}{III}, we show the subamplitudes $A^{(2)}$ and $A^{(4)}$ for the three-pion scattering with respect to the momentum $p$.
We can compare in size the leading-order and the next-to-leading-order contributions together with the constituents of the latter put together in several groups.
The endpoints of the plotted lines ($p=0.1$\,GeV) are consistent with the values shown in Table~\hyperlink{tab:I}{I}.
There we can see that significant cancellations take place.
In particular, some of the contributions related to the triangle integrals are sizable but cancel against each other to a negligible total contribution to the subamplitude.
The dominant contribution then stems from the polynomial parts ($A_\pi$, $A_L$, $A_l$) and the pieces that can be expressed in terms of the one-loop two-point functions $(A_J^{(1)}$, $A_J^{(2)}$).
We can also compare the 1PI part to the 1PR one.
\newpage
\noindent
%
\vspace{2mm}
\hrule\vspace{0.5mm}\hrule
\vspace{2mm}
{\noindent\strut\small
\sc Figure \hypertarget{fig:III}{III}.\ {\rm
    The scattering of three pions in the kinematic configuration of Eq.~\eqref{eq:kinematics}.
    We plot the LO result $A^{(2)}(p_1,p_2,\dots,p_6)$ of Eq.~\eqref{eq:defA2} and the NLO result $A^{(4)}(p_1,p_2,\dots,p_6)$ of Eq.~\eqref{eq:defA4}.
    Moreover, we show several groups of the individual constituents of $A^{(4)}$.
}
\strut
}
\begin{center}
    \includegraphics[width=0.99\columnwidth]{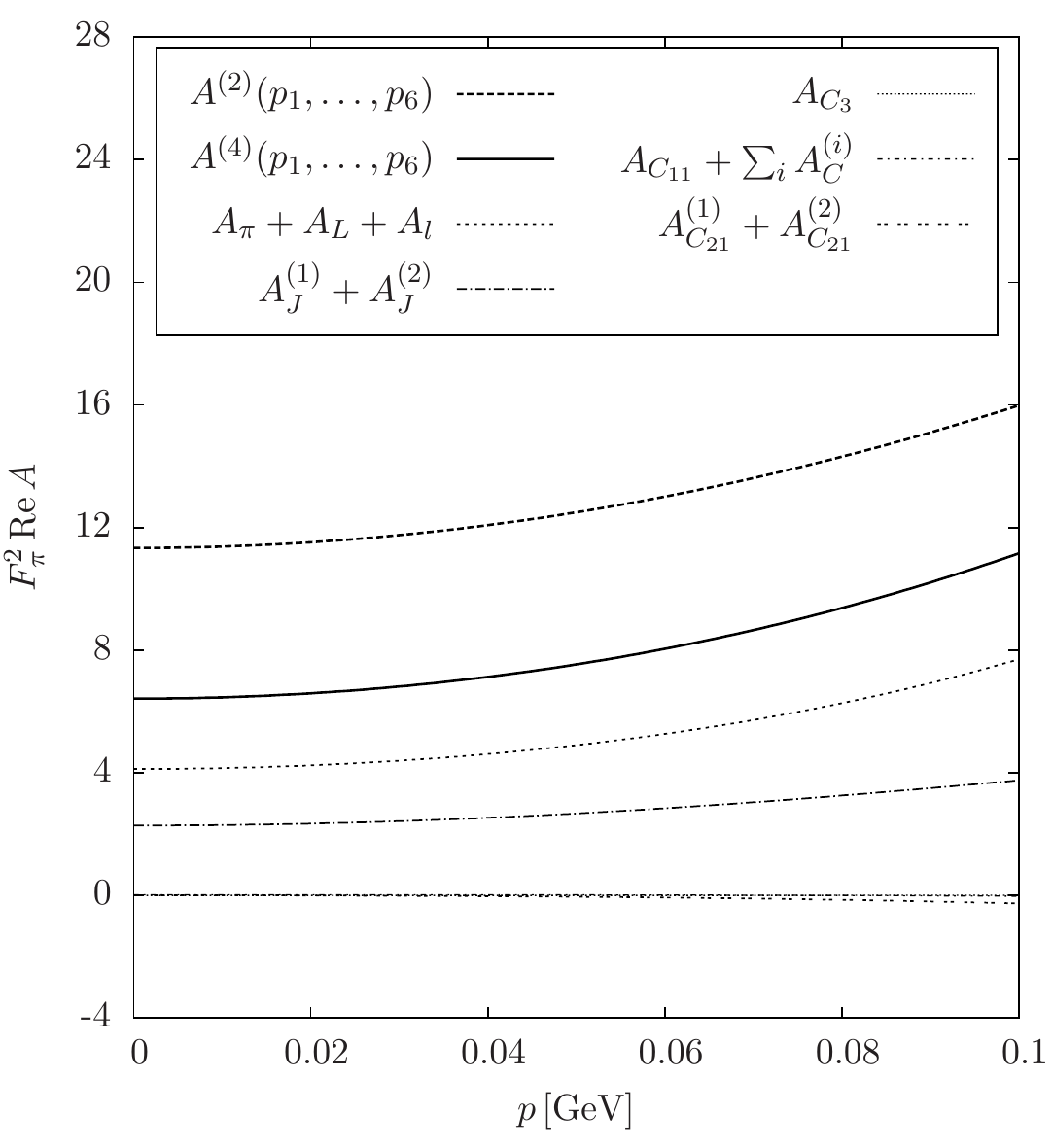}
    \label{fig:plotA}
    \vspace{2mm}
\end{center}
%
\vspace{2mm}
\hrule\vspace{0.5mm}\hrule
\vspace{2mm}
{\noindent\strut\small
\sc Table \hypertarget{tab:I}{I}.\ {\rm
    The 1PR and 1PI parts of the three-pion scattering amplitude, using four-momenta from Eq.~\eqref{eq:kinematics}, with $p=0.1\,\text{GeV}$.
    The amplitudes are all taken in the flavor-stripped form analogous to Eq.~\eqref{eq:A6piP15}.
    As in Eq.~\eqref{eq:defA4}, $A^{(4)}$ is the sum of the results we quote in the bottom part of the table.
    To obtain dimensionless quantities, the amplitudes are multiplied by a fitting power of $F_\pi$.
}
\strut
}
\vspace{1mm}
\begin{center}
\small{\renewcommand{\arraystretch}{1.4}
\renewcommand{\tabcolsep}{1.1pc}
    \begin{tabular}{c|r||c|r}
    \toprule
    \multicolumn{4}{c}{$F_\pi^2\times\operatorname{Re}A$} \\
    \midrule
        $A^{(4\pi)}_{6\pi}$ (\text{LO}) & \,$-319.00$ & \,$A^{(2)}$\, & 15.99 \\
        $A^{(4\pi)}_{6\pi}$ (\text{NLO}) & $-28.54$ & \,$A^{(4)}$\, & 11.16 \\
    \midrule\midrule
    \multicolumn{4}{c}{$F_\pi^2\times\operatorname{Re}A/F_\pi^6$} \\
    \midrule
        $A_{C_3}$  & 0.002 & $A_{J}^{(1)}$  & 1.917 \\
        $A_{C_{21}}^{(1)}$  & $-0.948$ & $A_{J}^{(2)}$  & 1.835 \\
        $A_{C_{21}}^{(2)}$  & 0.682 & $A_\pi$ & \,$-2.488$ \\
        $A_{C_{11}}$  & 0.090 & $A_L$ & 8.985 \\
        $A_{C}^{(1)}$  & $-0.026$ & $A_l$ & 1.209 \\
        $A_{C}^{(2)}$  & $0.890$ \\
        $A_{C}^{(3)}$  & $-0.984$ \\
        \bottomrule
    \end{tabular}
}
    \label{tab:numerics}
\end{center}
%
\noindent
We see that, at least in this kinematical setting, the LO of the six-pion subamplitude is roughly half the size of the NLO contribution of its pole counterpart.
On the other hand, the NLO contribution of the six-pion subamplitude is rather sizable compared to LO.
However, this is still acceptable in view of the whole six-pion amplitude, i.e.\ considering the dominance of the 1PR part.
Moreover, it is important to realize that at the three-pion threshold we happen to be at the edge of the applicability of ChPT.

Finally, we can find very simple analytical expressions in the limit $p\to0$:
\begin{align}
&\hphantom{\operatorname{Re}\,\,}F_\pi^2A^{(2)}\big|_{p\to0}
=5\,\frac{M_\pi^2}{F_\pi^2}\,,\\
\begin{split}
&F_\pi^2\operatorname{Re}A^{(4)}\big|_{p\to0}
=\frac{M_\pi^4}{F_\pi^4}\bigg\{
(-33+22N)\kappa
-\frac{1}{9}\kappa\\
&-\frac16(14+75N)L
+(16l_1^\text{r}+56l_2^\text{r}+6l_3^\text{r}+20l_4^\text{r})
\bigg\}\,.\notag
\end{split}
\label{eq:p0limit}
\end{align}
The contribution of the triangle integrals alone amounts to only $\frac\kappa2(9N-26)$ and is thus negligible for $N=3$.

\section{Summary}

We presented the NLO result for the four-pion and, most importantly, six-pion amplitudes, calculated in the massive O($N+1$)/O(N) nonlinear sigma model, the relevant Lagrangian of which shown in Eq.~\eqref{eq:L} leads to results consistent with two-flavor ChPT.
Our main result is the six-pion amplitude, which we split into 1PR and 1PI parts.
The 1PR part in Eq.~\eqref{eq:A6pi4pi} employs the form for the four-pion amplitude \eqref{eq:A4pisub_NLO} generalizing (beyond $N=3$) the results given in Refs.~\cite{Bijnens:1995yn,Bijnens:1997vq}.
The 1PI part \eqref{eq:A6piP15} of the six-pion amplitude --- represented by the six-pion subamplitude $A^{(4)}$ from Eq.~\eqref{eq:defA4} --- can be written in terms of a large number of subparts, each of which satisfies the expected permutation symmetries.
Due to a nontrivial but suitable choice of the symmetrical basis for the tensor triangle one-loop integrals, the final expressions can be written fairly compactly and can be found in Ref.~\cite{Bijnens:2021hpq}.
Numerically, the NLO correction is sizable with respect to the LO of the six-pion subamplitude, which is, however, suppressed compared to the 1PR part.

\section*{Acknowledgments}

This work is supported  in  part  by the Swedish Research Council grants contract numbers 2016-05996 and 2019-03779.

\end{multicols}

\medline

\begin{multicols}{2}
%

\begingroup
\begin{thebibliography}{10}

\bibitem{Weinberg:1978kz}
S.~Weinberg, ``{Phenomenological Lagrangians},''
  \href{http://dx.doi.org/10.1016/0378-4371(79)90223-1}{{\em Physica A}
  {\bfseries 96} no.~1-2, (1979) 327--340}.

\bibitem{Gasser:1983yg}
J.~Gasser and H.~Leutwyler, ``{Chiral Perturbation Theory to One Loop},''
  \href{http://dx.doi.org/10.1016/0003-4916(84)90242-2}{{\em Annals Phys.}
  {\bfseries 158} (1984) 142}.

\bibitem{Osborn:1969ku}
H.~Osborn, ``{Implications of Adler zeros for multipion processes},''
  \href{http://dx.doi.org/10.1007/BF02755724}{{\em Lett. Nuovo Cim.} {\bfseries
  2} (1969) 717--723}.

\bibitem{Low:2019ynd}
I.~Low and Z.~Yin, ``{Soft Bootstrap and Effective Field Theories},''
  \href{http://dx.doi.org/10.1007/JHEP11(2019)078}{{\em JHEP} {\bfseries 11}
  (2019) 078}.

\bibitem{Bijnens:2019eze}
J.~Bijnens, K.~Kampf, and M.~Sj\"o, ``{Higher-order tree-level amplitudes in
  the nonlinear sigma model},''
  \href{http://dx.doi.org/10.1007/JHEP11(2019)074}{{\em JHEP} {\bfseries 11}
  (2019) 074}. [Erratum: \href{http://dx.doi.org/10.1007/JHEP03(2021)066}{{\em JHEP} {\bfseries 03}
  (2021) 066}].

\bibitem{Mai:2018djl}
M.~Mai and M.~Doring, ``{Finite-Volume Spectrum of $\pi^+\pi^+$ and
  $\pi^+\pi^+\pi^+$ Systems},''
  \href{http://dx.doi.org/10.1103/PhysRevLett.122.062503}{{\em Phys. Rev.
  Lett.} {\bfseries 122} no.~6, (2019) 062503}.

\bibitem{Blanton:2019vdk}
T.~D. Blanton, F.~Romero-L\'opez, and S.~R. Sharpe, ``{$I=3$ Three-Pion
  Scattering Amplitude from Lattice QCD},''
  \href{http://dx.doi.org/10.1103/PhysRevLett.124.032001}{{\em Phys. Rev.
  Lett.} {\bfseries 124} no.~3, (2020) 032001}.

\bibitem{Mai:2019fba}
M.~Mai, M.~D\"oring, C.~Culver, and A.~Alexandru, ``{Three-body unitarity
  versus finite-volume $\pi^+\pi^+\pi^+$ spectrum from lattice QCD},''
  \href{http://dx.doi.org/10.1103/PhysRevD.101.054510}{{\em Phys. Rev. D}
  {\bfseries 101} no.~5, (2020) 054510}.

\bibitem{Culver:2019vvu}
C.~Culver, M.~Mai, R.~Brett, A.~Alexandru, and M.~D\"oring, ``{Three pion
  spectrum in the $I=3$ channel from lattice QCD},''
  \href{http://dx.doi.org/10.1103/PhysRevD.101.114507}{{\em Phys. Rev. D}
  {\bfseries 101} no.~11, (2020) 114507}.

\bibitem{Fischer:2020jzp}
M.~Fischer, B.~Kostrzewa, L.~Liu, F.~Romero-L\'opez, M.~Ueding, and C.~Urbach,
  ``{Scattering of two and three physical pions at maximal isospin from lattice
  QCD},'' \href{http://dx.doi.org/10.1140/epjc/s10052-021-09206-5}{{\em Eur.
  Phys. J. C} {\bfseries 81} no.~5, (2021) 436}.

\bibitem{Hansen:2020otl}
{\bfseries Hadron Spectrum} Collaboration, M.~T. Hansen, R.~A. Brice\~no, R.~G.
  Edwards, C.~E. Thomas, and D.~J. Wilson, ``{Energy-Dependent $\pi^+ \pi^+
  \pi^+$ Scattering Amplitude from QCD},''
  \href{http://dx.doi.org/10.1103/PhysRevLett.126.012001}{{\em Phys. Rev.
  Lett.} {\bfseries 126} (2021) 012001}.

\bibitem{Brett:2021wyd}
R.~Brett, C.~Culver, M.~Mai, A.~Alexandru, M.~D\"oring, and F.~X. Lee,
  ``{Three-body interactions from the finite-volume QCD spectrum},''
  \href{http://dx.doi.org/10.1103/PhysRevD.104.014501}{{\em Phys. Rev. D}
  {\bfseries 104} no.~1, (2021) 014501}.

\bibitem{Blanton:2021llb}
T.~D. Blanton, A.~D. Hanlon, B.~H\"orz, C.~Morningstar, F.~Romero-L\'opez, and
  S.~R. Sharpe, ``{Interactions of two and three mesons including higher
  partial waves from lattice QCD},''
  \href{http://dx.doi.org/10.1007/JHEP10(2021)023}{{\em JHEP} {\bfseries 10}
  (2021) 023}.

\bibitem{Bijnens:2021hpq}
J.~Bijnens and T.~Husek, ``{Six-pion amplitude},''
  \href{http://dx.doi.org/10.1103/PhysRevD.104.054046}{{\em Phys. Rev. D}
  {\bfseries 104} no.~5, (2021) 054046}.

\bibitem{Dobado:1994fd}
A.~Dobado and J.~Morales, ``{Pion mass effects in the large N limit of chiral
  perturbation theory},''
  \href{http://dx.doi.org/10.1103/PhysRevD.52.2878}{{\em Phys. Rev. D}
  {\bfseries 52} (1995) 2878--2890}.

\bibitem{Bijnens:2009zi}
J.~Bijnens and L.~Carloni, ``{Leading Logarithms in the Massive $O(N)$
  Nonlinear Sigma Model},''
  \href{http://dx.doi.org/10.1016/j.nuclphysb.2009.10.028}{{\em Nucl. Phys. B}
  {\bfseries 827} (2010) 237--255}.

\bibitem{Bijnens:2010xg}
J.~Bijnens and L.~Carloni, ``{The Massive $O(N)$ Non-linear Sigma Model at High
  Orders},'' \href{http://dx.doi.org/10.1016/j.nuclphysb.2010.09.019}{{\em
  Nucl. Phys. B} {\bfseries 843} (2011) 55--83}.

\bibitem{Bijnens:1995yn}
J.~Bijnens, G.~Colangelo, G.~Ecker, J.~Gasser, and M.~E. Sainio, ``{Elastic
  $\pi\pi$ scattering to two loops},''
  \href{http://dx.doi.org/10.1016/0370-2693(96)00165-7}{{\em Phys. Lett. B}
  {\bfseries 374} (1996) 210--216}.

\bibitem{Bijnens:1997vq}
J.~Bijnens, G.~Colangelo, G.~Ecker, J.~Gasser, and M.~E. Sainio, ``{Pion-pion
  scattering at low energy},''
  \href{http://dx.doi.org/10.1016/S0550-3213(97)00621-4}{{\em Nucl. Phys. B}
  {\bfseries 508} (1997) 263--310}. [Erratum:
  \href{https://doi.org/10.1016/S0550-3213(98)00127-8}{{\em Nucl. Phys. B}
  {\bfseries 517} (1998) 639--639}].

\bibitem{Bijnens:2014lea}
J.~Bijnens and G.~Ecker, ``{Mesonic low-energy constants},''
  \href{http://dx.doi.org/10.1146/annurev-nucl-102313-025528}{{\em Ann. Rev.
  Nucl. Part. Sci.} {\bfseries 64} (2014) 149--174}.

\bibitem{Colangelo:2001df}
G.~Colangelo, J.~Gasser, and H.~Leutwyler, ``{$\pi \pi$ scattering},''
  \href{http://dx.doi.org/10.1016/S0550-3213(01)00147-X}{{\em Nucl. Phys. B}
  {\bfseries 603} (2001) 125--179}.

\bibitem{Aoki:2016frl}
S.~Aoki {\em et al.}, ``{Review of lattice results concerning low-energy
  particle physics},''
  \href{http://dx.doi.org/10.1140/epjc/s10052-016-4509-7}{{\em Eur. Phys. J. C}
  {\bfseries 77} no.~2, (2017) 112}.

\end{thebibliography}
\endgroup
%
\providecommand{\href}[2]{#2}

\end{multicols}

\end{document}